\documentclass[a4paper,11pt]{article}
\pdfoutput=1 

\usepackage{color}   
\usepackage[margin=1.5cm]{geometry}
\usepackage{amsmath,amssymb}

\begin{document}

\title{\bf Prospects for a dominantly microwave-diagnosed 
magnetically confined fusion reactor}

\author{Francesco A. Volpe\\
  \em Dept of Applied Physics and Applied Mathematics\\ 
  \em Columbia University, New York, NY 10027, USA\\
fvolpe@columbia.edu}




\maketitle

\abstract{Compared to present experiments, tokamak and stellarator 
reactors will be subject 
to higher heat loads, sputtering, erosion and subsequent coating,  
tritium retention, higher neutron fluxes, and a number of radiation effects. 
Additionally, neutral beam penetration in tokamak reactors 
will only be limited to the plasma edge. As a result, several optical, 
beam-based and magnetic diagnostics of today's plasmas might not be applicable 
to tomorrow's reactors, but the present discussion suggests that reactors 
could largely rely on microwave diagnostics, including techniques based on mode 
conversions and Collective Thomson Scattering.}

\flushbottom

\section{Introduction}
The international ITER tokamak \cite{ITER07} is 
expected to achieve a fusion gain (ratio of fusion power to input heating 
power) $Q$=10 and to demonstrate the scientific feasibility of magnetic 
confinement fusion (MCF).

The Fusion Nuclear Science Facility (FNSF) \cite{FNSF,Kessel1,Kessel2} is a 
tokamak or spherical tokamak  
advocated in the United States, with emphasis on nuclear issues and 
on the integration of the first wall, blanket, shield, vacuum vessel and 
divertor in the nuclear fusion environment. 

DEMO \cite{DEMOZohm} is the common name for the fusion device 
-most likely a tokamak 
or a stellarator- expected to demonstrate the technological feasibility of MCF. 

ITER, FNSF, DEMO and a magnetic confinement 
fusion power plant will operate longer and longer plasma discharges and will 
generate larger and larger amounts of fusion power, in absolute terms as well 
as normalized to input power. Correspondingly, as they are all 
expected to be based on D-T reactions, they will generate larger and larger 
fluences of 14.1 MeV neutrons, resulting in increased material damage, as 
summarized in Table \ref{tab1}. 
First walls, blankets and coolants will also be hotter, as hot 
as about 600$^\circ$ C in a reactor.  
This poses special requirements on materials \cite{NatMat} and diagnostics 
\cite{Donne04,Hartfuss06,Donne07,Vayakis08,Weller,
Donne,RBoivin,Dolan,Todd,Orsitto}.

\begin{table}[t]
\centering
\caption{\label{tab1} Maximum damage, in displacements per atom (dpa), and 
typical plasma discharge duration expected in future 
MCF devices\cite{Kessel1,Kessel2}.}
\smallskip
\begin{tabular}{|lllll|}
\hline
& 
ITER &
FNSF &
DEMO & 
Power Plant 
\\
\hline
Max damage &
3 dpa & 
37-74 dpa &
100-150 dpa & 
$>$150 dpa 
\\
Plasma discharge duration & 
500-3000 s & 
1-15 days &
15-365 days &
$>$365 days  
\\
\hline
\end{tabular}
\end{table}

MCF reactors will not be physics experiments and, 
apart from initial scenario optimizations and periodic calibrations, they will 
not require several engineering or physics parameters to be scanned.   
Instead, they will be steadily operated in a smaller parameter space.   
Yet, precisely for this reason, plasma parameters will need to be 
continuously diagnosed with sufficient temporal and spatial resolution to 
prevent undesired instabilities as well to maintain optimal conditions 
(density, temperature, rotation, field errors, plasma shape, position and 
others) for confinement, fusion reactivity etc.  
For these reasons, it is important to develop reactor-grade diagnostics and 
ensure all relevant measurements. In decreasing order of importance, these 
are: measurements for machine 
protection and basic control, measurements for advanced control, and those 
for performance evaluation \cite{Donne04,Donne07,ReNeW}. 

At the same time, all measurements will have to be compatible with the 
harsh fusion reactor environment. That is, they will have to face the issues 
listed in Sec.\ref{SecChall} of the present paper, distilled from the vast  
literature on the subject 
\cite{Donne04,Hartfuss06,Donne07,Vayakis08,Weller,
Donne,RBoivin,Dolan,Todd,Orsitto}. 
Optical and inductive diagnostics will be particularly 
susceptible to such issues, and neutral beams will only be available 
in the outer part of the plasma.  
Possible countermeasures include shielding the diagnostics 
and investigating new materials (Sec.\ref{SecProtMat}) 
or relying as extensively as possible on microwave diagnostics that, 
despite their own issues \cite{GConwayIAEA}-\cite{Silva}, are 
more resilient to the reactor environment (Sec.\ref{SecRobustMicrow}). 

With this preamble, the present paper discusses {\em for the first time}  
the question of whether it will be possible to measure all or nearly 
all quantities of interest in a reactor by means of microwaves. 
The answer to this question is encouraging: 
as discussed in Sec.\ref{SecDirect} and summarized in Table \ref{tab2} of the 
present paper, more than two thirds of all the relevant  
observables listed in Refs.\cite{Donne04,Donne07,ReNeW} are {\em directly} 
measurable by either well-proven 
or less established but promising microwave techniques  
-based for example on Collective Thomson Scattering, Electron Bernstein Waves  
or mode conversions. 
Emphasis is laid on the physical principles enabling the measurements. 

Another goal of the present article is to identify research needs and 
{\em future} directions: 
as summarized in Table 3, there are still several quantities that, at present, 
can only be measured {\em indirectly} by microwaves (Sec.\ref{SecIndirect}), 
or not at all (Sec.\ref{SecResNeeds}), but it would be highly 
desirable to directly measure them by means of microwaves. 
It goes beyond the scope of the present paper to review 
established microwave diagnostics and {\em past} results. 
For that purpose, the reader is redirected to a book \cite{Hartfuss} 
and review papers \cite{Mase,Luhmann,KraemerFl}.

\section{Diagnostic challenges in a MCF reactor}\label{SecChall}
Diagnostics in a reactor will face \cite{Donne,RBoivin,Orsitto,Dolan}:
\begin{itemize}

    \item High heat loads.

    \item High fluences 
    of energetic particles, leading to sputtering and erosion. 
    Among others, this leads to the coating of optics by impurities and 
    polymers \cite{Mirr09,Mirr12,Mirr13,Mirr14,Mirr15,Mirr16a}.

    \item High fluences of 14.1 MeV neutrons, carrying  
      80\% of the power produced in D-T fusion reactions. 
      At hundreds or thousands of MW, and divided by the wall surface, this 
      implies exposure of the walls to about 1 MW/m$^2$. Significant amounts 
      of 2.45 MeV D-D neutrons will also be produced. This results in 
      ``nuclear heating'' \cite{FederYoussef} 
      of the walls and underlying materials, as well as 
      in transmutations \cite{Gilbert}. The latter affect semiconductors 
      by transmuting the bulk material or the dopant 
      into another element with different 
      donor or acceptor properties \cite{Bolshakova,IAEATransmut}. 

      \item Radiation-induced electromotive force due to 
        Compton and photo-electrons \cite{Zinkle1,Zinkle2,Cables}. 
        This electromotive force can lead to 
        spurious voltages along as well as across mineral-insulated 
        coaxial cables. Note that other insulations (for example, plastic) 
        would be radiation-damaged too quickly. Inductive diagnostics will 
        suffer from such issues. 

      \item Radiation-induced degradation or alteration of material 
        properties. This includes:

        \begin{itemize}

          \item The introduction of finite conductivity in 
            insulators \cite{Zinkle3,nRIC,Vayakis,Hartfuss06}, 
            affecting inductive diagnostics of magnetic field, 
            bolometers and pressure gauges.
            
          \item The darkening of refractive optics 
            \cite{GMcKee}. 
            This is a volume effect due to radiation, distinct from 
            superficial sputtering and erosion effects. 

          \item Radiation-induced luminescence (radioluminescence) 
            in windows and optical fibers 
            \cite{Gorshkov,Costley01,Vayakis,Hartfuss06}.

          \item Damage of solid-state components and detectors 
            \cite{Hall,CCD} as used in 
            infrared, visible, ultraviolet, X-ray and neutron cameras, which 
            can be contained by proper doping and radiation modification 
            \cite{Bolshakova}.
        \end{itemize}

      \item Tritium retention issues \cite{Skinner}, 
        for example in spectrometers \cite{TritiumSpectr}. 
        
      \item Vibrations and large electromagnetic forces, 
        introducing statistical and systematic errors in diagnostic alignment. 

      \item Thermoelectric effects, whether radiation-induced 
        \cite{RITES1,RITES2} or not \cite{Moreau}, resulting in a 
        thermally induced electromotive force.

      \item Reduced neutral-beam penetration \cite{Kanev,ITER-NBI,Rej} 
        due to the large minor 
        radius, especially in a tokamak reactor. This might affect 
        beam-based diagnostics, restrict their applicability to 
        the plasma edge and prevent the core from being diagnosed. 
        In addition, the signal-to-noise ratio is expected to degrade 
        as a result of the high density $n$ and large minor radius $a$. 
        This is because the signal grows like $n$, but Bremsstrahlung 
        grows like $n^2a$ \cite{Rej}. Diagnostics affected include:  
        Beam Emission Spectroscopy (BES), 
        measuring electron density fluctuations \cite{BES}; 
        Charge Exchange Recombination (CER), measuring ion temperatures 
        and flow velocities \cite{CER}; 
        Motional Stark Effect (MSE), measuring the magnetic field 
        rotational transform \cite{MSE-Wolf}. 
        All three diagnostics are highly relevant 
        to tokamak confinement and stability, but the last two 
        are less needed in stellarators. 

        \item
          Tokamak reactors will have additional needs and challenges compared 
          with stellarator reactors, due to disruptions and 
          runaway electron beams.
\end{itemize}

If follows from this list that several diagnostics commonly 
used in present tokamaks might have issues or be unavailable in a reactor. 
Beam penetration issues will affect BES, CER, MSE and Lithium 
beam spectroscopy. Sputtering and coating of windows and mirrors, and 
volumetric radiation damage to optical materials will affect 
optical diagnostics such as Thomson Scattering, 
Survey Poor Resolution Extended Domain (SPRED) spectroscopy, 
visible Bremsstrahlung and other spectrometers, filterscopes, 
phase contrast imaging, 
infrared, visible and ultraviolet cameras, including fast cameras. 
Radiation damage, including neutron damage, will affect 
inductive magnetic diagnostics, bolometers and pressure gauges. 
This and thermal damage would affect diagnostics in close proximity to the 
plasma and first wall, such as tile-embedded thermocouples, or even in direct 
contact with the plasma edge, such as Langmuir probes.

\section{Diagnostic protection and new materials}\label{SecProtMat}

One could think of several strategies to counteract the issues above.

One possible strategy is obviously to protect existing diagnostics from 
neutrons and radiation in general 
(by means of shielding), and from exposure to plasma, energetic 
particles and stray electromagnetic waves (e.g.~by means of shutters, 
to limit their utilization to when strictly necessary). Cooling 
can protect from high heat-loads and heat-transients. 

Even with these precautions, however, survival in DEMO 
would be limited to about three months for magnetics, one week for bolometry 
and few hours for pressure gauges and VUV windows, and this is only 
accounting for radiation effects in DEMO \cite{ReNeW}; survival in a plant, and 
with all effects included, would be even shorter. 
Also note that laser- and microwave-based 
techniques for in-situ cleaning of windows, lenses and mirrors 
\cite{Mirr12,Mirr14,Mirr15,Mirr16a}, 
while useful and indispensable in ITER, do not prevent volumetric 
damage by neutrons, especially in DEMO or a power plant. 

The next level of protection is 
to identify new materials for existing techniques. New materials 
are being sought, which are more resilient to radiation effects and 
transmutations. Platinum, in this sense, is a good replacement 
for gold \cite{Giannone,Peterson}. 
Single-crystal \cite{Voitsenya} and liquid mirrors \cite{LiqMirr} 
are being considered, because radiation-harder than regular mirrors. 

Inevitably, though, lenses and mirrors will be damaged, which implies 
that their use should be minimized, and direct lines  
of sight should be adopted as much as possible.

\section{Advantages and disadvantages of microwave diagnostics}   
\label{SecRobustMicrow}

Another approach is complementary and possibly alternative to the 
diagnostic protection and material innovations just mentioned in 
Sec.\ref{SecProtMat}. 
The idea is to rely on long wavelengths 
such as far infrared, THz, millimeter waves and microwaves: in proportion 
with the longer wavelength, diagnostics are less sensitive 
to the polishing and finish of windows, mirrors and lenses, 
and less sensitive to sputtering and erosion damage. The ``grooves'' and 
rough features resulting from sputtering, erosion and other damage 
are typically much smaller, both in width and depth, than the wavelength 
and skin-depth $\delta=\sqrt{2\rho / \omega \mu}$. Here $\rho$ is the bulk 
resistivity and $\mu=\mu_0 \mu_r$ the magnetic permeability. For frequencies 
$\omega/2\pi=$1-100 GHz, relative permeability of the medium 
$\mu_r\simeq$1 and typical metal resistivities, the skin depth 
amounts to 0.2-7 $\mu$m.

Another effect is coating: material sputtered elsewhere can deposit on 
microwave optics. However, when such material is dielectric, 
its effect on microwave attenuation, for typical coating  
thicknesses of few $\mu$m or less, refractive index $N$=1-2 and loss tangent 
of $10^{-5}$-$10^{-2}$ \cite{Goldsmith}, is negligible. 
Reflectivity can deteriorate due to dielectric material deposition, 
but mostly in the utraviolet and visible range; 
the reflectivity of dielectric materials in the 
microwave range is comparatively much higher \cite{Adachi}. 
A metalic film, on the other hand, can have a significant shielding effect and 
make a window or a lens opaque, 
in so far as it is thicker than the skin depth and ``holes'' in the coating 
are not bigger than the wavelength. 
The effect on microwave mirrors, instead, is probably negligible: the 
reflectivity of the metallic film deposited might be lower than 
that of the metal underneath, but only by few percents \cite{Adachi}, 
which are easily accounted for in a recalibration. 

The adoption of even longer wavelengths, 
thus lower frequencies ($\lesssim$500 MHz) is probably to be avoided, 
because antennas and inductive coils face other issues related 
to cable damage (see Sec.\ref{SecChall}).  

That being said, reactor-grade microwave diagnostics will not be 
immune from practical issues either \cite{GConwayIAEA}.  
For instance, antennas will need to be carefully designed and positioned, 
and antennas' arrays carefully configured \cite{GConwayIAEA} with the aid of 
ray-tracing, beam-tracing \cite{Stegmeir2} and full-wave codes 
\cite{Kramer,SilvaFDTD} as to collect acceptable signals despite refraction and 
vertical movements of the plasma. 
Concerns on limited access led to proposals of adopting a single 
spherical mirror \cite{Liu14} or Rowland circle optics \cite{Liu15} for 
microwave imaging.  
Windows and other in-vessel and in-port components will have to withstand 
microwave stray radiation \cite{GConwayStray,Oosterbeek}, 
electromagnetic loads, neutron activation, and  
meet maintenance, remote handling and safety requirements 
\cite{Sirinelli,Udintsev}. They will also have to 
be properly designed to enable diagnostic calibration 
\cite{GConwayITER,GConwayCalib}.  
Transmission lines will be longer than in present devices and will 
require high mode purity and low losses and reflections \cite{Bigelow,Hanson}. 
Finally the magnetic field will be higher, thus 
Electron Cyclotron Emission (ECE), Collective Thomson Scattering and other 
diagnostics will require sources \cite{sourceCTS} and receivers \cite{recECE} 
to operate at higher frequencies.

Yet, these issues are considered solved or solvable in the conceptual 
and detailed designs of the ITER ECE radiometer \cite{Rowan,Austin,Taylor}, 
reflectometers \cite{VayakisRefl} 
and other microwave diagnostics \cite{Donne07}. All things considered, even in 
a reactor, these issues are less challenging than those 
to be faced by non-microwave diagnostics, listed above.    
Microwave signals might be difficult to interpret due for 
instance to the relativistic downshift of the emission frequency 
\cite{Bornatici,Silva}, but at least they will be available, unlike others, 
and they are expected to offer satisfactory signal-to-noise ratio, 
spatial and temporal resolution 
\cite{Rowan,Austin,Taylor,VayakisRefl,Donne07,Silva}.

\section{Important observables and their microwave and non-microwave 
diagnostics}\label{SecDiagn}
Refs.\cite{Donne04,Donne07,ReNeW} compiled a table of  
measurements needed in ITER and, with good approximation, in a reactor.  
Diagnostics were grouped by their purpose, in decreasing order of urgency: 
1a) machine protection and basic control, 1b) advanced control and 2) 
performance evaluation and physics.

That table is reproduced in Tables \ref{tab2} and \ref{tab3} here, with the 
additional classification of measurability by microwaves, which can 
be direct (Sec.\ref{SecDirect}), indirect (Sec.\ref{SecIndirect}) or 
pose a possible research need (Sec.\ref{SecResNeeds}).

\subsection{Direct measurements by microwaves}    \label{SecDirect} 
Microwave emission, interference, 
reflection, refraction and scattering provide information on the local or 
line-averaged dielectric tensor $\epsilon$, or selected components.
These contain information on the electron density $n_e$, 
temperature $T_e$, mean flow, and magnetic field 
\cite{Hartfuss,Mase,Luhmann,KraemerFl}.  

Some microwave measurements are well known and well established 
\cite{Hartfuss,Mase,Luhmann,KraemerFl}. These include 
$T_e$ profile measurements by Electron Cyclotron Emission (ECE) 
and $n_e$ measurements by interferometry. The latter are line-intergrated, 
but profiles can be inverted from multi-chord measurements. 
These diagnostics can also measure $n_e$ and $T_e$ fluctuations, 
thanks to their high temporal resolution or to related concepts such as 
Correlation ECE (CECE) \cite{Cima,White10,White12}.

Some microwave diagnostics can measure magnetic fields. Examples include 
polarimetry by Faraday rotation 
\cite{Bergerson} or Cotton-Mouton effect \cite{Fuchs}. 
Both techniques are line-integrated and both are sensitive to 
density and to a magnetic field component 
(transverse and parallel to the line of sight, respectively) \cite{Segre}. 
They can thus 
measure that field component, provided that the density profile is known 
otherwise. 

Here, however, it is more useful to discuss less common microwave 
techniques and discuss how they can fill the gaps to be left in a reactor by 
inductive, optical, atomic-beam diagnostics, and others. 

\begin{table}[t]
\centering
\caption{\label{tab2} 
Measurements needed in ITER and, with good approximation, in a 
reactor, adapted from Refs.\cite{Donne04,Donne07,ReNeW}. The observables are 
categorized by urgency and by the capability of microwave diagnostics to 
directly or indirectly measure them. See also Table \ref{tab3}.
}
\smallskip
\begin{tabular}{|p{20mm}|p{40mm}|p{40mm}|p{40mm}|}
\hline
& Group 1A
& Group 1B 
& Group 2 
\\
Microwave meas.~capability
& Meas.~for machine protection and basic control
& Measurements for advanced control
& Performance evaluation and physics 
\\
\hline
Direct (Secs.\ref{SecDirect} and \ref{SecScatt}) 
&
\begin{minipage}[t]{40mm}
\begin{flushleft}
  - Plasma shape and position, separatrix-wall gaps, gap between separatrices\\
  - Line-averaged electron density\\
  - Runaway electrons\\
\end{flushleft}
\end{minipage}
&
\begin{minipage}[t]{40mm}
\begin{flushleft}
  - Plasma rotation (toroidal and poloidal)\\ 
  - Electron temperature profile (core)\\
  - Electron density profile (core and edge)\\
  - Ionization front position in divertor\\
  - $n_e$ of divertor plasma\\
\end{flushleft}
\end{minipage}
&
\begin{minipage}[t]{40mm}
\begin{flushleft}
  - $T_e$ profile (edge)\\
  - $n_e$, $T_e$ profiles (X-point)\\
  - $T_e$ fluctuations\\
  - $n_e$ fluctuations\\
  - Edge turbulence
\end{flushleft}
\end{minipage}
\\
\hline
\begin{flushleft}
Direct, by CTS (Sec.\ref{SecCTS}) 
\end{flushleft}
&
\begin{minipage}[t]{40mm}
\begin{flushleft}
  - Impurity and D, T influx (divertor \& main plasma)\\
  - $Z_{eff}$ (line-averaged)\\
  - $n_T/n_D$ in plasma core
\end{flushleft}
\end{minipage}
&
\begin{minipage}[t]{40mm}
\begin{flushleft}
  - $\alpha$-source profile\\
  - Helium density profile (core)\\
  - Ion temperature profile (core)\\
  - $Z_{eff}$ profile\\
  - Helium density (divertor)\\
  - Impurity density profiles\\ 
  - $\alpha$-particle loss
\end{flushleft}
\end{minipage}
&
\begin{minipage}[t]{40mm}
\begin{flushleft}
  - $T_i$ in divertor\\
  - Confined $\alpha$ particles\\
  - $n_T/n_D/n_H$ (edge)\\
  - $n_T/n_D/n_H$ (divertor)\\
\end{flushleft}
\end{minipage}
\\
\hline
Direct, by mode conversions (Sec.\ref{SecMC})
&
  - Plasma current, $q_{95}$
&
\begin{minipage}[t]{40mm}
\begin{flushleft}
  - Current density profile ($q$-profile)\\
  - Low $m/n$ MHD activity\\
  - $T_e$ of divertor plasma
\end{flushleft}
\end{minipage}
&
  - TAE modes, fishbones
\\
\hline
Indirect (Sec.\ref{SecIndirect}) 
&
\begin{minipage}[t]{40mm}
\begin{flushleft}
  - Fusion power\\
  - $\beta_N=\beta_{tor}(aB/I_p)$\\
  - Disruption precursors (locked modes, $m$=2)\\
  - H/L mode indicator\\
  - ELMs
\end{flushleft}
\end{minipage}
&
- Sawteeth
&
- MHD activity in plasma core
\\
\hline
\end{tabular}
\end{table}

Starting with measurements needed for machine 
protection and basic control (group 1A in Refs.\cite{Donne04,Donne07,ReNeW} 
and Table \ref{tab2}), some of the 
observables ``at risk'' are plasma shape and position, separatrix-wall gaps 
and the gap $\delta_{sep}$ between the separatrices for the upper and lower 
null. Traditionally these quantities are reconstructed by EFIT \cite{EFIT} 
or other equilibrium reconstruction code on the basis of magnetic-probe and 
saddle-loop measurements. However, more recently the radial position of the 
plasma was measured reflectometrically, and adjusted in feedback with that 
measurement \cite{AUG,posITER}. The cutoff density and thus the 
microwave frequencies are lower than those used 
for density profile measurements. Several reflectometers operating at slightly 
different frequencies, as to not cross-talk with each other, could locate the 
separatrices at various poloidal locations, for 2D equilibrium reconstructions. 
With the addition of toroidally displaced reflectometers, 3D equilibria 
could be reconstructed as well. 

The ionization front position in the divertor can also be measured 
reflectometrically. 

Runaway electrons form tails in the electron distribution function, which 
can be measured by oblique 
\cite{Giruzzi,Preische96,Preische97,delaLuna,Simonetto} 
or vertical ECE \cite{Kato86,Kato87,Luce}. 

Additional plasma 
parameters can be directly measured by Collective Thomson Scattering, 
mode-conversion-based and other 
scattering diagnostic, as discussed in Secs.\ref{SecCTS}-\ref{SecScatt}. 

\subsubsection{Collective Thomson Scattering}\label{SecCTS}
Collective Thomson Scattering (CTS) \cite{Sitenko,Aamodt}  
is the scattering of electromagnetic waves off the 
electrons in the Debye spheres associated with ions. Thus, the scattered waves 
contain information on the ions,  
provided that the Salpeter parameter 
$\alpha=1/|{\bf k}_s|\lambda_D$ is larger than 1. Here ${\bf k}_s$ is the 
scattering wavevector and $\lambda_D$ the Debye length. Unless a 
back-scattering geometry is adopted, which has the drawback of 
poor spatial resolution, 
the criterion $\alpha >$1 translates in the requirement for low frequencies. 
Typical frequencies are 70-250 GHz, provided that the plasma is underdense 
to them. A CTS source must satisfy additional requirements of narrow 
spectrum, high power (as a consequence of the small scattering cross-section) 
and long pulses or continuous operation \cite{OrsiGiru,Lau}. The 
typical source of choice is the gyrotron, 
although the Cyclotron Autoresonance Maser recently regained attention 
\cite{sourceCTS}. 

The CTS spectrum mimics the 
ion distribution function \cite{BindslevPRL}, 
or a convolution of ion distribution functions if 
multiple species are present, due to impurities, to two fuel ions (D and T) 
and due to fusion $\alpha$'s. 

If the ions of the main species 
are Maxwellian, CTS can measure their temperature $T_i$ \cite{Suvorov}. 
It can also measure related quantities, such as flows in the 
direction of the scattering vector, simply resulting in a shift of the said 
distribution \cite{Castiglioni,Bindslev07}. 

The convolution is easier to separate if the species of interest have 
dramatically different ion velocity distributions, due for example to very 
different masses and/or energies. 
This implies that CTS can measure 
the slowdown of $\alpha$'s due to collisions, to radiations and 
to other effects \cite{Bindslev07,Salewski}, 
the concentration of high $Z$ impurities, and $Z_{eff}$ 
\cite{OrsittoRSI,Stejner}. 
In principle, for sufficiently high precision (requiring an intense 
source and a low-noise radiometer), CTS can also measure the 
ion ratio $n_T/n_D$  or $n_T/n_D/n_H$, and core density of He ashes 
\cite{Salewski,Stejner}.

All measurements are easily resolved in space, by crossing the incident 
microwave beam with several receiver beams. That is, profiles can be easily 
acquired of all the said quantities, and studied as a function of time, 
enabling transport studies of $\alpha$ particles and distinction between 
confined and lost $\alpha$'s. 

Perturbative, time-resolved experiments show promise for transport studies, 
e.g.~of D, T and impurity influx.

\subsubsection{Mode conversion based techniques}\label{SecMC}
The O-X mode conversion can be used to locally diagnose the field 
${\bf B}$ and related quantities such as the edge safety factor 
$q_{95}=aB_T/RB_p$. Here $a$ and $R$ are the plasma minor and 
major radius, and $B_T$ and $B_p$ are the toroidal and poloidal field. 
From $B_p$ one can determine the plasma current $I_p=\mu_0B_p/2\pi a$, 
typically measured by Rogowski coils. In fact, good part 
of the $q$ profile (or, equivalently, current profile) 
can be measured by mode conversion based techniques, 
offering a complement and possible 
replacement for the optical, beam-based MSE \cite{MSE-Wolf} and other 
non-microwave techniques \cite{TFR,Soltwisch,WolfJET,Gentle,Donne02}.  

The basic idea is that a 
special ${\bf B}$-dependent view makes the O and X-modes degenerate and not 
evanescent at the $n$-dependent O-mode cutoff layer. 
The angular map of conversion efficiency around that optimal 
direction also contains information on the local ${\bf B}$: 
the inclination of the conversion contours at various frequencies $f$ gives 
the inclination of field-lines at various radial locations. 

This was confirmed by 
two-dimensional scans \cite{Shevchenko,SpinMirr}, including rapid scans 
performed during a single discharge by means of a spinning mirror 
\cite{SpinMirr}, and eventually  
measurements simultaneously carried out with multiple 
sensors \cite{SFreethy13}. These initial demonstrations required 
the plasma to be overdense and emit Electron Bernstein (B) Waves (EBWs) 
that converted to the X-mode and eventually to 
the O-mode. 

The more recent proposal and simulation of 
oblique reflectometry imaging \cite{Meneghini} shows that 
the O-X conversion of an externally injected wave has advantages 
over the B-X-O conversion of internally emitted EBWs. 
The idea is now that, instead of a peak in transmissivity, the 
diagnostic characterizes a minimum in reflectivity, obtained for the same 
special, ${\bf B}$-dependent direction. This is more flexible and can 
be operated at arbitrary densities and fields, as it 
does not require the plasma to be an overdense EBW emitter. 
The signal-to-noise ratio is also superior, by adopting a sufficiently intense 
external reflectometric source. The spatial and temporal resolution are 
reflectometer-like (of the order of mm and sub-ms), more than sufficient 
for a reactor. 

As a consequence of measuring the ${\bf B}$ components along a fixed chord 
with good sensitivity, space- and time-resolution, the diagnostic 
is expected to be sensitive to oscillations of ${\bf B}$ associated 
with {\em rotating} modes in the plasma, such as tearing modes of 
low poloidal/toroidal mode numbers $m/n$, Toroidal Alfv\'{e}n Eigenmodes (TAEs) 
and fishbones. 

Future modeling might clarify whether this technique could also 
detect and characterize non-rotating 
magnetic islands and other modes in the plasma, 
for example from the deformation or displacement of 
the conversion efficiency contours, or 
field stochasticity, from fluctuations and distortions in the contours. 

The O-X conversion is sufficient for the technique just described. 
In principle, B-X-O converted EBW emission could also be of use in reactors, 
to measure $T_e$ in overdense plasmas not accessible to 
conventional ECE. It should be noted that  
at the very high field of a reactor, $B_T\ge$5T, 
the fundamental EC harmonic becomes inacessible at a very high density,  
$n_e\ge 2.4 \cdot 10^{20}$m$^{-3}$, which is unlikely in a tokamak. 
However, these high densities could 
be achieved in a stellarator high density H-mode \cite{McCormick}. 

Another mode conversion based technique of potential use in a reactor  
is EBW emission from the {\em underdense} edge and divertor region 
\cite{CostleyPriv}.  
Measurements of $T_e$ in this optically thin region might be difficult by 
conventional ECE. However, the plasma 
is optically thick to EBWs. Note that EBW emission and propagation do not 
require the plasma to be overdense (only the B-X-O conversion does).
The only requirement to couple EBWs with external electromagnetic waves is for 
the Upper Hybrid layer (but not necessarily the O-mode cutoff) 
to lie in the plasma. This condition is met by 
wave frequencies $\omega \approx \omega_{ce}$, where $\omega_{ce}$ is the 
electron cyclotron frequency in the divertor region. 
EBW emission could be extracted by B-(SX)-FX conversion \cite{Sugai}, where 
SX and FX refer to the slow and fast X-mode.

\subsubsection{Other scattering techniques}\label{SecScatt}
Microwave scattering also assures turbulent fluctuations of electron 
density $n_e$, through Doppler broadening of sharp lines (much narrower 
than 1 MHz) in the tens of GHz range of frequencies. 
This led to measurements of electron temperature gradient (ETG) turbulence 
\cite{Mazzucato}. 
The shift of the scattered spectrum relative to the incident line 
informs about the electron diamagnetic drift or, more generally, flows, 
including plasma rotation. 

Cross-Polarization Scattering \cite{CPS}, as the name suggests, 
examines scattering in which an incident O-mode (or X-mode) 
changes its polarization to X-mode (or O-mode). These changes of polarization 
are due to magnetic field {\em fluctuations} $\delta B$ in the plane 
perpendicular to ${\bf k}$. The actual modulus of ${\bf B}$, however, 
or its components, are unknown.

Another technique \cite{Vann16} aims at measuring the magnetic pitch angle 
based on the fact that turbulent structures are elongated in the pitch angle 
direction. In turn, such structures can be visualized by imaging 
scattered or reflected microwaves.

\subsection{Indirect measurements by microwaves}   \label{SecIndirect}
The evolution of the pedestal, Edge Localized Modes (ELMs), the distance from 
the peeling-ballooning stability boundary, the quiescent H-mode, I-mode 
and other pedestal physics can simply be studied by ECE, 
possibly combined with multi-chord interferometry. 

The normalized beta, $\beta_N=\beta_{tor}aB/I_p$, can be indirectly measured 
from the prescribed magnetic field $B$ and from the microwave measured 
minor radius $a$, plasma current $I_p$ and $\beta_{tor}$, which, in turn, is 
defined as the ratio between the measured kinetic pressure 
$k_B(n_eT_e+n_iT_i)$, where $k_B$ is the Boltzmann constant, and 
prescribed magnetic pressure $B_T^2/2\mu_0$. Similar considerations apply to 
the stored plasma energy and to the poloidal beta, $\beta_p$. 

The fusion power in a 
D-T reactor equals five times the energy associated with the $\alpha's$, 
measurable by CTS, divided by the energy confinement 
time, which can be perturbatively inferred from microwave-measured profiles 
of $n_e$, $n_i$, $T_e$, $T_i$. 

Finally, direct microwave measurements of magnetic field 
components, e.g.~associated with rotating modes, 
were discussed in Sec.\ref{SecMC}, but it should be added that 
MHD activity can also be measured indirectly, from the associated  
temperature fluctuations, which can be measured by ECE. 
Examples include disruption precursors such as the rotating $m$=2 mode, 
sawteeth, 
and core MHD, not accessible by oblique reflectometry imaging. The diagnosis 
of non-rotating (``locked'') MHD requires arrays of ECE radiometers, or ECE 
imaging.

\begin{table}[t]
\centering
\caption{\label{tab3} 
Measurements needed in ITER and, with good approximation, in a 
reactor, but currently not possible by microwave techniques 
(Sec.\ref{SecResNeeds}). Adapted from Refs.\cite{Donne04,Donne07,ReNeW}. 
See also Table \ref{tab2}.
}
\smallskip
\begin{tabular}{|p{20mm}|p{40mm}|p{40mm}|p{40mm}|}
\hline
& Group 1A
& Group 1B 
& Group 2 
\\
& Meas.~for machine protection and basic control
& Measurements for advanced control
& Performance evaluation and physics 
\\
\hline
Divertor
&
\begin{minipage}[t]{40mm}
\begin{flushleft}
  - Surface temperature (divertor \& upper plates)\\
  - Divertor detachment indicator\\
  - $j_{\rm sat}$, $n_e$, $T_e$ at divertor plate\\
\end{flushleft}
\end{minipage}
&
\begin{minipage}[t]{40mm}
\begin{flushleft}
  - Heat deposition profile (divertor)\\
  - Net erosion (divertor plate)\\
\end{flushleft}
\end{minipage}
&
\\
\hline
Gas
&
\begin{minipage}[t]{40mm}
\begin{flushleft}
  - Gas pressure (divertor \& duct)\\
  - Gas composition (divertor \& duct)\\
\end{flushleft}
\end{minipage}
&
\begin{minipage}[t]{40mm}
\begin{flushleft}
  - Neutral density between plasma and first wall\\
\end{flushleft}
\end{minipage}
&
\\
\hline
Radiation
&
- Radiated power (main plasma, X-point, divertor)
&
- Radiated power profile (core, X-point, divertor)
&
\\
\hline
Neutrons
&
&
\begin{minipage}[t]{40mm}
\begin{flushleft}
  - Neutron profile\\
  - Neutron fluence
\end{flushleft}
\end{minipage}
&
\\
\hline
Electro\-magnetic
&
\begin{minipage}[t]{40mm}
\begin{flushleft}
  - Loop voltage\\
  - Halo currents\\
\end{flushleft}
\end{minipage}
&
&
- Radial electric field and field fluctuations
\\
\hline
Other
&
\begin{minipage}[t]{40mm}
\begin{flushleft}
  - Surface temperature (first wall)\\
  - Dust
\end{flushleft}
\end{minipage}
&
&
\\
\hline
\end{tabular}
\end{table}

\subsection{Observables difficult to measure by microwaves, 
future research needs}    \label{SecResNeeds}
Some measurement objectives are not microwaves, ``by definition''. 
These include observables at different wavelengths, such as visible 
line-emission or X-ray Bremsstrahlung, or simply broadband-integrated 
measurements of radiation.  They also include other 
particles besides photons -mostly fusion neutrons. 

Measuring such observables by means of microwaves would require up- or 
down-conversion from different frequencies to microwaves. 
Unfortunately this is not a very common need; conversion to visible light 
(for example from X-rays, by means of a CsI (Tl) scintillator, 
or from infrared, by upconversion nanoparticles) is in much higher demand. 

Neutron measurements would require scintillators that are sensitive to neutrons 
and emit in the microwave or far infrared range and, at the same time, are 
resilient to radiation damage, 
probably amorphous (plastic or liquid), and resilient to transmutations.  

Measurements of gas pressure and composition represent 
another diagnostic gap in a reactor (Table \ref{tab3}). It will be 
important to develop alternatives to conventional pressure gauges and 
multi gas analyzers (MGAs), bearing in mind that infrared and 
microwave spectroscopy are sensitive to vibrational and rotational 
transitions, respectively, in {\em molecules}. Atomic spectroscopy 
requires visible and ultraviolet light. 

The need for innovative gas diagnostics is particularly strong in the 
divertor region. Other innovations are needed for the thermal and 
surface characterization of the divertor plates, as well as to measure $n_e$, 
$T_e$ and the ion-saturation current-density $j_{sat}$ in the nearby plasma 
(Table \ref{tab3}).

The challenges to be faced by magnetic diagnostics 
were described in Sec.\ref{SecChall}. 
Innovations might be needed in diagnosing 
the loop voltage, typically measured by flux loops, and halo currents 
during disruptions, typically measured by Rogowski and segmented 
Rogowski coils \cite{SGerhardt}. 

Dust presence and dust dynamics is also of concern, e.g.~for fear of 
disruptions \cite{Lazzaro}. 
These and other quantities, currently not measurable by microwave techniques, 
are listed in Table \ref{tab3}.

\section{Summary and conclusions}\label{}
Magnetic confinement fusion reactors will not be physics experiments and, 
apart from initial scenario optimizations and periodic calibrations, 
for the most part they will be steadily operated at constant, optimal  
parameters for confinement, stability and fusion reactivity.    
Yet, precisely for this reason, plasma parameters will need to be 
continuously diagnosed with sufficient temporal and spatial resolution to 
maintain such optimal conditions.   

Unfortunately, however, today's 
magnetic, optical and beam-based diagnostics 
will face various challenges in the harsh reactor environment. 
Microwave and direct-line-of-sight diagnostics, on the other hand, 
are more robust. 
For these reasons, in parallel with ongoing research on new materials, 
radiation hardening, neutron shielding and in-situ cleaning of 
existing diagnostics, it is estimated that new and existing microwave 
techniques could diagnose more than half of the relevant observables 
(Tables \ref{tab2} and \ref{tab3}). 
Among others, a more extensive use of 
reflectometry is advocated, to replace magnetics in measuring the plasma 
shape and position. Collective Thomson Scattering proved successfull at 
diagnosing ion parameters that are 
normally measured by optical, beam-based diagnostics. 
Finally, recent proposals could enable internal, local measurements of 
magnetic field based on mode-conversion oblique reflectometry imaging. 

On the other hand diagnostic innovations, possibly based on microwaves, 
might be needed in the areas of divertor, gas, neutron, radiation and some 
electromagnetic measurements.



\end{document}